# Nonlinear sub-switching regime of magnetization dynamics in photo-magnetic garnets

A. Frej, I. Razdolski, A. Maziewski, and A. Stupakiewicz

Faculty of Physics, University of Bialystok, 1L Ciolkowskiego, 15-245 Bialystok, Poland

**Abstract.** We analyze, both experimentally and numerically, the nonlinear regime of the photo-induced coherent magnetization dynamics in cobalt-doped yttrium iron garnet films. Photo-magnetic excitation with femtosecond laser pulses reveals a strongly nonlinear response of the spin subsystem with a significant increase of the effective Gilbert damping. By varying both laser fluence and the external magnetic field, we show that this nonlinearity originates in the anharmonicity of the magnetic energy landscape. We numerically map the parameter workspace for the nonlinear photo-induced spin dynamics below the photo-magnetic switching threshold. Corroborated by numerical simulations of the Landau-Lifshitz-Gilbert equation, our results highlight the key role of the cubic symmetry of the magnetic subsystem in reaching the nonlinear spin precession regime. These findings expand the fundamental understanding of laser-induced nonlinear spin dynamics as well as facilitate the development of applied photo-magnetism.

1. **INTRODUCTION**

Recently, a plethora of fundamental mechanisms for magnetization dynamics induced by external stimuli at ultrashort time scales has been actively discussed [1-5]. The main interest is not only in the excitation of spin precession but in the switching of magnetization between multiple stable states, as it opens up rich possibilities for non-volatile magnetic data storage technology. One of the most intriguing examples is the phenomenon of ultrafast switching of magnetization with laser pulses. Energy-efficient, non-thermal mechanisms of laser-induced magnetization switching require a theoretical understanding of coherent magnetization dynamics in a strongly non-equilibrium environment [6]. This quasiperiodic motion of magnetization is often modeled as an oscillator where the key parameters, such as frequency and damping, are considered within the framework of the Landau-Lifshitz-Gilbert (LLG) equation [1, 7]. Although it is inherently designed to describe small-angle spin precession within the linear approximation, there are attempts to extend this formalism into the nonlinear regime where the precession parameters become angle-dependent [8]. This is particularly important in light of the discovery of the so-called precessional switching, where magnetization, having been impulsively driven out of equilibrium, ends its precessional motion in a different minimum of the potential energy [6, 9-11]. Obviously, such magnetization trajectories are characterized by very large precession angles (usually on the order of tens of degrees). It is, however, generally believed that the magnetization excursion from the equilibrium of about 10-20 degrees is already sufficient for the violation of the linear LLG approach [12, 13]. Thus, an intermediate regime under the switching stimulus threshold exists, taking a large area in the phase space and presenting an intriguing challenge in understanding fundamental spin dynamics.

An impulsive optical stimulus often results in a thermal excitation mechanism, inducing concomitant temperature variations, which can impact the parameters of spin precession [14-16]. This highlights the special role of the non-thermal optical mechanisms of switching [17-19]. Among those, we outline photo-magnetic excitation, which has been recently demonstrated in dielectric Co-doped YIG (YIG:Co) films [6, 11]. There, laser photons at a wavelength of 1300 nm resonantly excite the $^5E \rightarrow\ ^5T_2$ electron transitions in Co-ions, resulting in an emerging photo-induced magnetic anisotropy and thus in a highly efficient excitation of the magnetic subsystem [6]. This photo-induced effective anisotropy field features a nearly instantaneous rise time (within the femtosecond pump laser pulse duration), shifting the equilibrium direction for the magnetization and thus triggering its large-amplitude precession. In the sub-switching regime (at excitation strengths just below the switching threshold), the frequency of the photo-induced magnetization precession has been shown to depend on the excitation wavelength [20]. However, nonlinearities in magnetization dynamics in the sub-switching regime have not yet been described in detail, and the underlying mechanism for the frequency variations is not understood.

In this work, we systematically examine the intermediate sub-switching regime characterized by large angles of magnetization precession and the nonlinear response of the spin system to photo-magnetic excitations. We show a strong increase of the effective Gilbert damping at elevated laser-induced excitation levels and quantify its nonlinearity within the existing phenomenological formalism [8]. We further map the nonlinear regime in the phase space formed by the effective photo-induced anisotropy field and the external magnetic field.

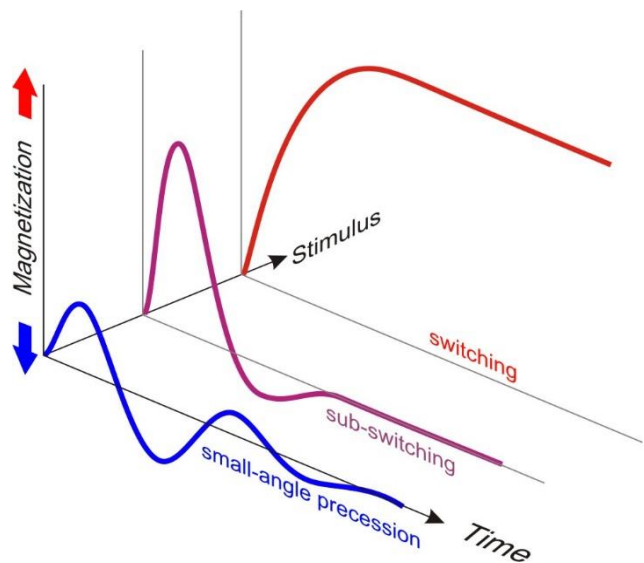

**Fig. 1.** Sketch of magnetization dynamics at various stimulus levels. Owing to the highly nonlinear magnetization dynamics in the switching regime, the nonlinearity onset manifests in the sub-switching regime too.

This paper is organized in the following order: in the first part, we describe the details of the experiment for laser-induced large-amplitude magnetization precession. Next, we present the experimental results, followed by the fitting analysis. Then, we complement our findings with the results of numerical simulation of the photo-magnetic spin dynamics. Afterward, we discuss the workspace of parameters for the sub-switching regime of laser-induced magnetization precession. The paper ends with conclusions.

## 2. EXPERIMENTAL DETAILS

The experiments were done on a 7.5 µm-thick YIG:Co film with a composition of $Y_2CaFe_{3.9}Co_{0.1}GeO_{12}$. The Fe ions at the tetrahedral and octahedral sites are replaced by Co-ions [21]. The sample was grown by liquid-phase epitaxy on a 400 µm-thick gadolinium gallium garnet (GGG) substrate. It exhibits eight possible magnetization states along the garnet's cubic cell diagonals due to its cubic magnetocrystalline anisotropy ($K_1 = -8.4 \times 10^3 \; erg/cm^3$) dominating the energy landscape over the uniaxial anisotropy ($K_u = -2.5 \times 10^3 \; erg/cm^3$). Owing to the 4° miscut, additional in-plane anisotropy is introduced, tilting the magnetization axes and resulting in slightly lower energy of half of the magnetization states in comparison to the others. In the absence of the external magnetic field, the equilibrium magnetic state corresponds to the magnetization in the domains close to the <111>-type directions in YIG:Co film. Measurements of the Gilbert damping $\alpha$ using the ferromagnetic resonance technique resulted in $\alpha \approx 0.2$. This relatively high damping is inextricably linked to the Co dopants [22-24].

The nonlinearity of an oscillator is usually addressed by varying the intensity of the stimulus and comparing the response of the system under study. Here, we investigated the nonlinear magnetization dynamics by varying the optical pump fluence and, thus, the strength of the photo-magnetic effective field driving the magnetization out of the equilibrium. We performed systematic studies in various magnetic states of YIG:Co governed by the magnitude of the external magnetic fields. The magnetic field $H_\perp$ was applied perpendicular to the sample plane and in-plane magnetic field $H$ was applied along the [110] direction of the YIG:Co crystal by means of an electromagnet. Owing to the introduced miscut, the studied YIG:Co exhibits four magnetic domains at $H = 0$ [25]. The large jump at an in-plane magnetic field close to zero shows the magnetization switching in the domain structures between four magnetic phases. The optical spot size in this experiment was around 100 µm while the size of smaller domains was around 5 µm, resulting in the spatial averaging of the domains in the measurements. This behavior of magnetic domains was discussed and visualized in detail by magneto-optical Faraday effect in our previous papers [6, 25]. With an increase of the magnetic field up to around $H = 0.4$ kOe, larger and smaller domains are formed due to the domain wall motion, eventually resulting in a formation of a single domain in a noncollinear state. Upon further increase, the magnetization rotates towards the direction of the applied field until a collinear state with in-plane magnetization orientation is reached at about 2 kOe (see Fig. 2).

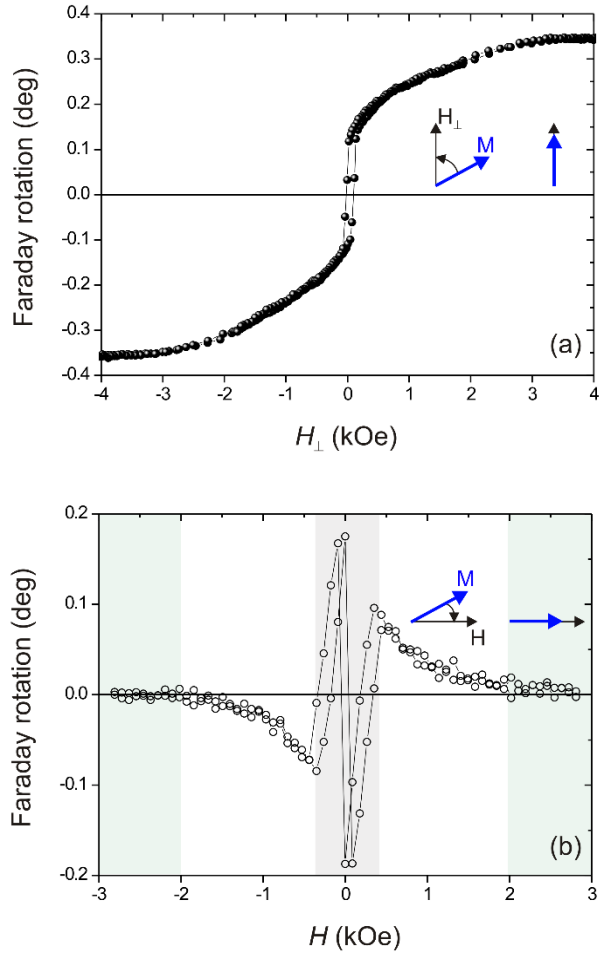

**Fig. 2.** Magnetization reversal using static magneto-optical Faraday effect under perpendicular $H_\perp$ (a) and in-plane $H$ (b) magnetic fields. The grey area indicates the magnetization switching in magnetic domain structure [25]. The green area shows the saturation range with a collinear state of magnetization.

Dynamic nonlinearities in the magnetic response were studied employing the pump-probe technique relying on the optical excitation of the spin precession in YIG:Co film. The pumping laser pulse at 1300 nm, with a duration of 50 fs and a repetition rate of 500 Hz, induced spin dynamics through the photo-magnetic mechanism [6]. The transient Faraday rotation of the weak probe beam at 625 nm was used to monitor the dynamics of the out-of-plane magnetization component $M_z$. The diameter of the pump spot was around 140 μm, while the probe beam was focused within the pump spot with a size of around 50 μm. The fluence of the pump beam was varied in the range of 0.2 – 6.5 mJ/cm$^2$, below the switching threshold of about 39 mJ/cm$^2$ [20]. At 1300 nm pump wavelength, the optical absorption in our garnet is about 12%. An estimation of the temperature increase $\Delta T$ due to the heat load for the laser fluence of 6.5 mJ/cm$^2$ results in $\Delta T <1$ K (see Methods of Ref. 6). The polarization of both beams was linear and set along the [100] crystallographic direction in YIG:Co for the pump and the [010] direction for the probe pulse. The experiments were done at room temperature. At each magnetic field, we performed a series of laser fluence-dependent pump-probe experiments measuring the transients of an oscillating magnetization component normal to the sample plane. We then used a phenomenological damped oscillator response function to fit the experimental data and retrieve the fit parameters such as amplitude, frequency, lifetime and effective damping. In what follows, we analyze the obtained nonlinearities in the response of the magnetic system and employ numerical simulations to reproduce the experimental findings.

### 3. RESULTS
### A. Time-resolved photo-magnetic dynamics

In order to determine the characteristics of the photo-magnetic precession, we carried out time-resolved measurements of a transient Faraday rotation $\Delta\theta_F$ in YIG:Co film. Fig. 3(a-d) exemplifies a few typical datasets obtained for four various pump fluences (between 1.7 and 6.5 mJ/cm$^2$) in magnetic fields of various strengths. A general trend demonstrating a decrease of the precession amplitude and an increase of its frequency is seen upon the magnetic field increase. To get further insights into the magnetization dynamics, these datasets were fitted with a damped sine function on top of a non-oscillatory, exponentially decaying background:

$$\Delta\theta_F(\Delta t) = A_F \sin(2\pi f \Delta t + \phi) \exp\left(-\frac{\Delta t}{\tau_1}\right) + B \exp\left(-\frac{\Delta t}{\tau_2}\right), \qquad (1)$$

where $\Delta t$ is pump and probe time difference, $A_F$ is the amplitude, $f$ is the frequency, $\phi$ is the phase, $\tau_1$ is the decay time of precession, and $\tau_2$ is the decay time of the background with an amplitude $B$.

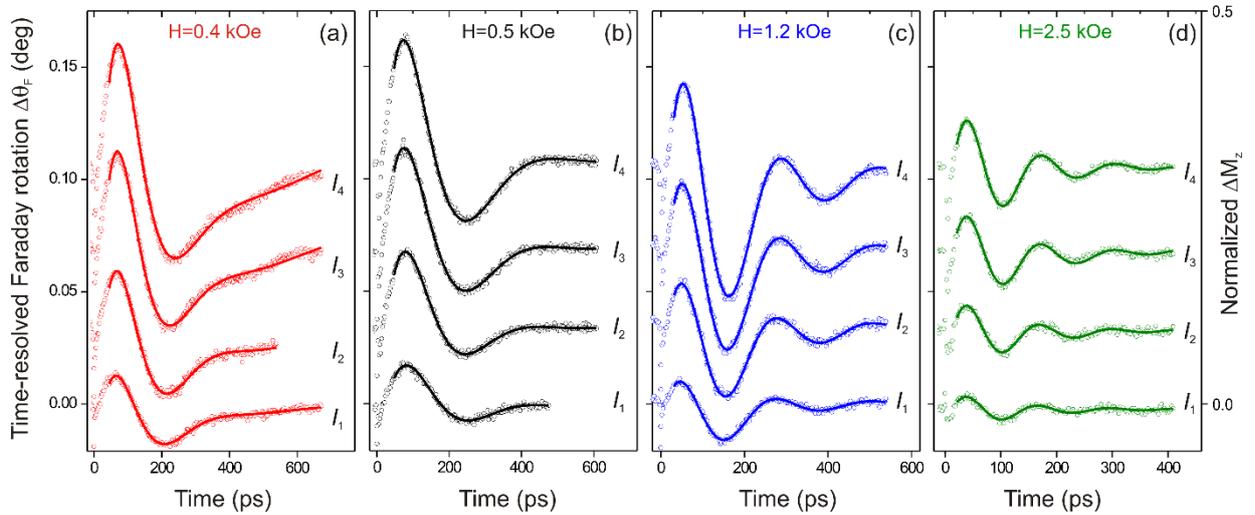

**Fig. 3.** Time-resolved Faraday rotation at different magnetic fields $H$ (a-d) and laser fluences ($I_1$-$I_4$ correspond to 1.7, 3.2, 5.0, and 6.5 mJ/cm$^2$, respectively). The normalized $\Delta M_z$ on the vertical axis is defined as $\Delta\theta_F/\theta_{max}$, where $\theta_{max}$ is obtained for saturation magnetization rotation at $H_\perp$ (see. Fig. 2a). The curves are offset vertically without rescaling. The solid lines are fittings with the damped sine function (Eq. 1).

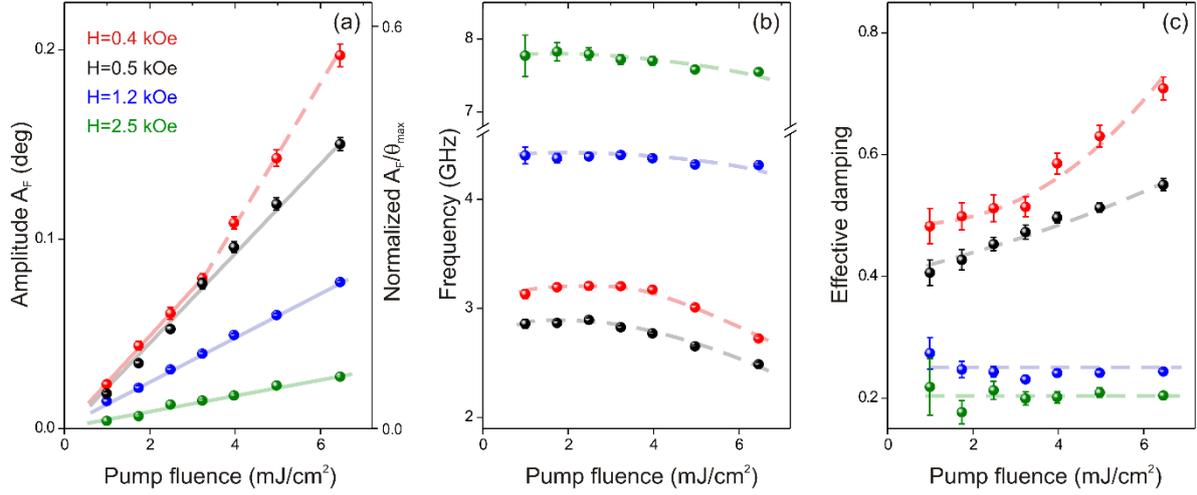

**Fig. 4.** Photo-magnetic precession parameters as a function of pump fluence in different external magnetic field $H$: a) amplitude of the Faraday rotation $A_F$, b) frequency of the precession, and c) effective damping. Different colors correspond to different external magnetic fields. The solid lines are the linear fits where applicable, while the dashed lines are the visual guides. Some of the error bars are smaller than the data point symbols.

At low applied fields $H < 1$ kOe, where the photo-magnetic anisotropy field ($H_L$) contribution to the total effective magnetic field is the strongest, the largest magnetization precession amplitude is observed. Figure 4 shows the most important parameters of the magnetization precession, that is, amplitude, frequency and effective damping (Fig. 4a-c). The latter is obtained from the frequency and the lifetime as $(2\pi f \tau_1)^{-1}$. Although the amplitude dependence on the pump fluence is mostly linear, the other two parameters exhibit a more complicated dependence, which is indicative of the noticeable nonlinearity in the magnetic system. In particular, at $H = 0.4$ and $0.5$ kOe, we observed an increase in the effective damping with laser fluence, resulting in a faster decay of the magnetic precession. This is further corroborated by the frequency decrease seen in Fig. 4b. It is seen that the behavior of the magnetic subsystem is noticeably dissimilar at low (below 1 kOe) and high (above 2 kOe) magnetic fields. At higher magnetic fields $H > 1$ kOe we were unable to observe nonlinear magnetization response at pump fluences up to 10 mJ/cm². This is indicative of a significant difference in the dynamic response in the collinear and noncollinear states of the magnetic subsystem.

### 4. Nonlinear precession of magnetization in anisotropic cubic crystals

The data shown in Fig. 4c clearly indicates the nonlinearity in the magnetic response manifesting in the increase of the effective damping with the excitation (laser) fluence. Previously, similar behavior was found in a number of metallic systems [26-29] and quickly attributed to laser heating. Interestingly, Chen et al. [30] found a decrease of the effective damping with laser fluence in FePt, while invoking the temperature dependence of magnetic inhomogeneities to explain the results. There, the impact of magnetic inhomogeneity-driven damping contribution exhibits a similar response to laser heating and an increase in the static magnetic field. A more complicated mechanism relying on the temperature-dependent competition between the surface and bulk anisotropy contributions and resulting in the modification of the effective anisotropy field has been demonstrated in ultrathin Co/Pt bilayers [31, 32].

Nonlinear spin dynamics is a rapidly developing subfield enjoying rich prospects for ultrafast spintronics [33]. Importantly, all those works featured thermal excitation of magnetization dynamics in metallic, strongly absorptive systems. In stark contrast, we argue that the mechanism in the Co-doped YIG studied here is essentially non-thermal. This negligible temperature change Δ$T$ is unable to induce significant variations of the parameters in the magnetic system of YIG:Co ($T_N$=450 K), thus ruling out the nonlinearity mechanism discussed above. Rather, we note the work by Müller et al. [34], where the non-thermal nonlinear regime of magnetization dynamics in $CrO_2$ at high laser fluences was ascribed to the spin-wave instabilities at large precession amplitudes [35]. We also note the recently debated and physically rich mechanisms of magnetic nonlinearities, such as spin inertia [36-39] and relativistic effects [40, 41]. Yet, we argue that in our case of a cubic magnetic anisotropy-dominated energy landscape, a much simpler explanation for the nonlinear spin dynamics can be suggested. In particular, we attribute the amplitude-dependent effective damping to the anharmonicity of the potential well for magnetization.

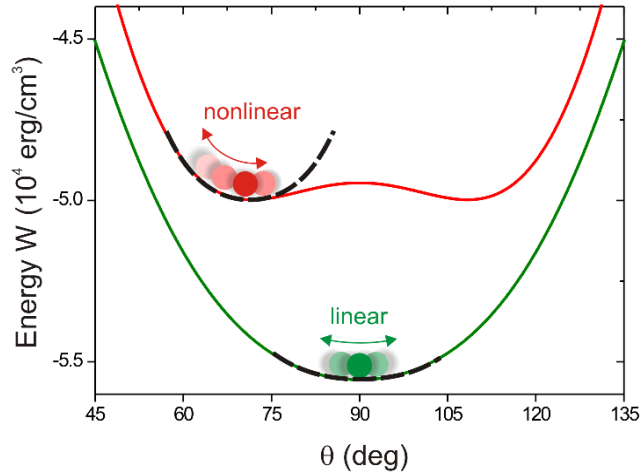

**Fig. 5.** Energy landscape as a function of the polar angle $\theta$ ($\varphi = 45°$) in the linear ($H = 2.5$ kOe, green) and nonlinear ($H = 0.4$ kOe, red) precession regimes. The dashed lines are the parabolic fits in the vicinity of the minima. $\theta$ is the polar angle of magnetization orientation measured from the normal to the sample plane along the [001] axis in YIG:Co.

We performed numerical calculations of the energy density landscape $W(\theta, \varphi)$:

$$W(\theta, \varphi) = W_c + W_u + W_d + W_z \qquad (2)$$

taking into account the following terms in the free energy of the system: the Zeeman energy $W_z = -\mathbf{M} \cdot \mathbf{H}$, demagnetizing field term $W_d = -2\pi M_s^2 \sin^2 \theta$, cubic $W_c = K_1 \cdot (\sin^4 \theta \sin^2 \varphi \cos^2 \theta + \sin^2 \theta \cos^2 \theta \cos^2 \varphi + \sin^2 \theta \cos^2 \theta \sin^2 \varphi)$ and uniaxial anisotropy $W_u = K_u \sin^2 \theta$ ($\theta$ and $\varphi$ are the polar and azimuthal angles, respectively). In the calculations, we assume $K_1 = -9 \cdot 10^3$ erg/cm³, $K_u = -3 \cdot 10^3$ erg/cm³, and $M_s$ is the saturation magnetization of 7.2 Oe [25]. Then, following [8] and [42], we calculate the precession frequency $f$ and the effective damping $\alpha_{eff}$:

$$f = \frac{\gamma}{2\pi M_s \sin\theta} \sqrt{\frac{\delta^2 W}{\delta\theta^2}\frac{\delta^2 W}{\delta\varphi^2} - \left(\frac{\delta^2 W}{\delta\theta\delta\varphi}\right)^2}, \quad (3)$$

$$\alpha_{eff} = \frac{\alpha_0 \gamma \left(\frac{\delta^2 W}{\delta\theta^2} + \frac{\delta^2 W}{\delta\varphi^2}\sin^{-2}\theta\right)}{8\pi^2 f M_s}, \quad (4)$$

where the $\gamma$ is gyromagnetic ratio, and $\alpha_0$ is the Gilbert damping in YIG:Co [23, 24]. In Fig. 5, we only show the total energy as a function of the polar angle $\theta$, to illustrate the anharmonicity of the potential at small external in-plane magnetic fields. Experimental data and calculations of the energy $W(\theta, \varphi)$ have been published in Refs. [25, 43]. There, it is seen that at relatively small external magnetic fields canting the magnetic state, the proximity of a neighboring energy minimum (to the right) effectively modifies the potential well for the corresponding oscillator (on the left), introducing an anharmonicity. On the other hand, at sufficiently large magnetic fields, which, owing to the Zeeman energy term, modify the potential such that a single minimum emerges (shown in Fig. 5 in green), no nonlinearity is expected. This is also in line with the decreasing impact of the cubic symmetry in the magnetic system, which is responsible for the anharmonicity of the energy potential.

To get yet another calculated quantity that can be compared to the experiment, we introduced the photo-magnetically induced effective anisotropy term $K_L$. This contribution depends on the laser fluence *I* through the effective light-induced field $H_L \propto I$ as:

$$K_L = -2H_L M_s \cos^2\theta \quad (5)$$

The presence of this term displaces the equilibrium for net magnetization. The equilibrium directions can be obtained by minimizing the total energy with and without the photo-magnetic anisotropy term. Then, knowing the angle between the perturbed and unperturbed equilibrium directions for the magnetization, we calculated the precession amplitude $A$. We note the difference between the amplitudes $A_F$, which refers to the Faraday rotation of the probe beam, and $A$ standing for the opening angle of magnetization precession. Although both are measured in degrees, their meaning is different.

Having repeated this for a few levels of optical excitation, we obtained a linear slope of the amplitude vs excitation strength dependence. Figure 6(a-c) illustrates the amplitude, frequency, and (linear) effective damping as a function of the external magnetic field. The agreement between the calculated parameters and those obtained from fitting the experimental data is an impressive indication of the validity of our total energy approach. Further, the linear effective damping value of $\alpha \approx 0.2$ obtained in the limit of strong fields, is in good agreement with the values known for our Co-doped YIG from previous works [6, 24]. In principle, the effective damping in garnets can increase towards lower magnetic fields. Conventionally attributed to the extrinsic damping contributions, this behavior has been observed in rare-earth iron garnets before as well and ascribed to the generation of the backward volume spin wave mode by ultrashort laser pulses [44]. It is worth noting that there is no nonlinearity phenomenologically embedded in the approach given above.

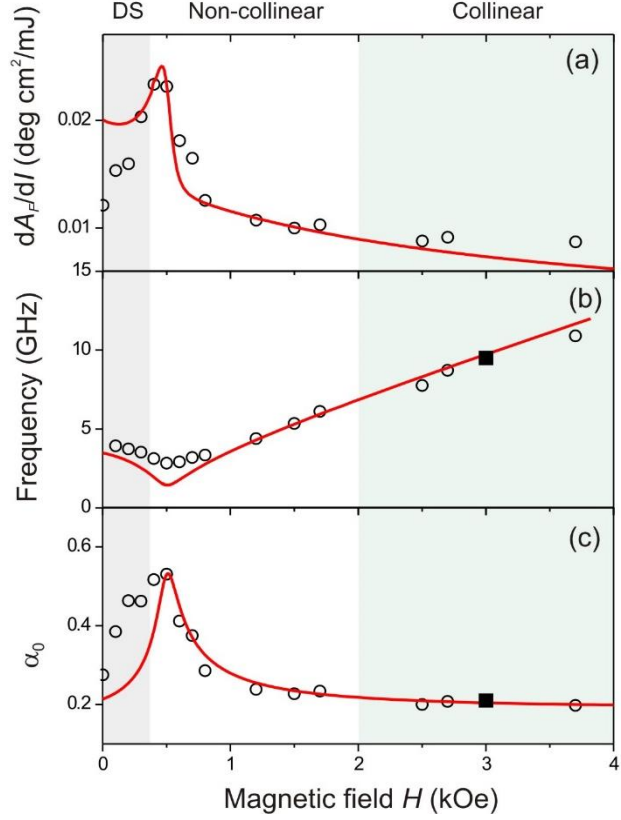

**Fig. 6.** Photo-magnetic precession parameters at various magnetic fields: amplitude (a), frequency (b), and (linear) effective damping (c). The points are from the experimental data, the solid lines are calculated as described in the text. The dark rectangular points are obtained in the FMR experiments. The grey shaded area indicates the presence of a domain state (DS). The green shaded area shows the magnetization saturation state.

Yet, the data presented in Fig. 4c indicates the persistent nonlinear behavior of the effective damping. To clarify the role of the potential anharmonicity, we fitted the potentials $W(\theta, \varphi)$ using a parabolic function with an anharmonic term:

$$W(x) = W_0 + k[(x - x_0)^2 + \beta_x(x - x_0)^4] \qquad (6)$$

Here $x = \theta$ or $\varphi$, and $\beta_x$ is the anharmonicity parameter. We calculated it independently for $\theta$ and $\varphi$ for each dataset of $W(\theta, \varphi)$ obtained at different values of the external magnetic field $H$ by fitting the total energy with Eq. (6) in the vicinity of the energy minimum (Fig. 5). This anharmonicity should be examined on equal footing with the nonlinear damping contribution. To quantify the latter, we follow the approach by Tiberkevich & Slavin [8] and analyze the effective damping dependencies on the precession amplitude by means of fitting a second-order polynomial to them:

$$\alpha = \alpha_0 + \alpha_2 A^2. \qquad (7)$$

The examples of the fit curves are shown in Fig. 7a, demonstrating a good quality of the fit within a certain range of the amplitudes $A$ (below 45°). It should, however, be noted that the model in Ref. [8] has been developed for the in-plane magnetic anisotropy, and thus its applicability for our case is limited. This is the reason why we do not go beyond the amplitude dependence of the effective damping and do not analyze the frequency dependence on $A$ in the limit of strong effective fields. We note that the amplitude $A$, the opening angle of the precession, should be understood as a mathematical parameter only, and not as a true excursion angle of magnetization obtained in the real experimental conditions. There, large effective Gilbert damping values and a short decay time of the photo-magnetic anisotropy preclude the excursion of magnetization from its equilibrium to reach these $A$ values.

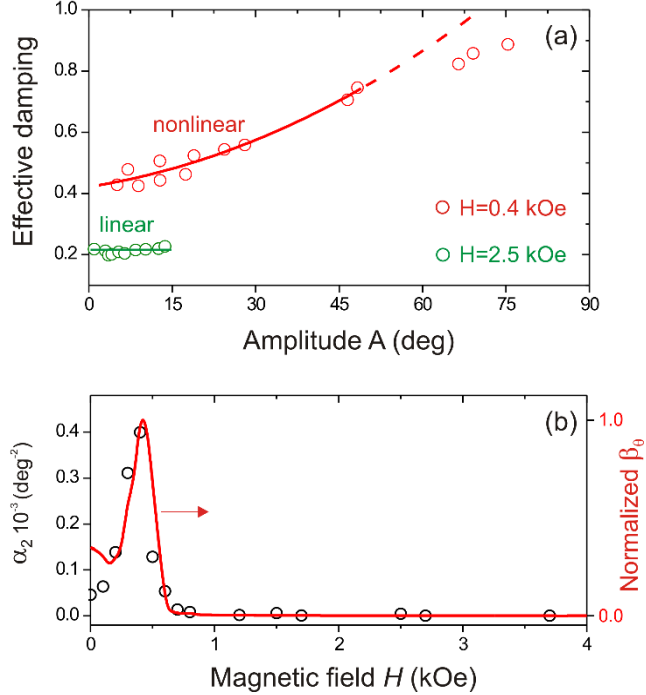

**Fig. 7.** a) Effective damping in the linear and nonlinear precession regimes of the precession amplitude $A$. The lines are the second-order polynomial fits with Eq. (7). b) Magnetic field dependence of the nonlinearity parameters: nonlinear damping coefficient $\alpha_2$ (points, obtained from experiments) and the $W(\theta)$ potential anharmonicity normalized $\beta_\theta$ (red line, calculated).

We note that the anharmonicity parameter $\beta_x$ calculated for the $W(\theta)$ profiles was found to be a few orders of magnitude larger than that obtained for $W(\varphi)$. This difference in the anharmonicity justifies our earlier decision to focus on the shape of $W(\theta)$ potential only (cf. Fig. 5). This means that the potential for magnetization in the azimuthal plane is much closer to the parabolic shape and much larger amplitudes of the magnetization precession are required for it to start manifesting nonlinearities in dynamics. As such, we only consider the anharmonicity $\beta_x$ originating in the $W(\theta)$ potential energy. In Fig. 7b, we compare the $\beta_\theta$ (red line) and $\alpha_2$ (points) dependencies on the external in-plane magnetic field. It is seen that its general shape is very similar, corroborating our assumption that the potential anharmonicity is the main driving force behind the observed nonlinearity. We argue that thanks to the cubic magnetic anisotropy in YIG:Co film, the potential anharmonicity-related mechanism of nonlinearity allows for reaching the nonlinear regime at moderate excitation levels.

### 5. Simulations of laser-induced magnetization dynamics

To further prove that the observed nonlinearities in magnetization dynamics do not require introducing additional inertial or relativistic terms [33], we complemented our experimental findings with numerical simulations of the LLG equation:

$$\frac{d\mathbf{M}}{dt} = -\gamma[\mathbf{M} \times \mathbf{H}_{\text{eff}}(t)] + \frac{\alpha}{M_s}\left(\mathbf{M} \times \frac{d\mathbf{M}}{dt}\right), \qquad (8)$$

where $H_{eff}$ is the effective field derived from Eq. (2) as:

$$\mathbf{H}_{\text{eff}}(t) = -\frac{\partial W_A}{\partial \mathbf{M}} + \mathbf{H}_{\text{L}}(t), \qquad (9)$$

We employed the simulation model from Ref. [11] and added a term corresponding to the external magnetic field $H$. Calculations performed for a broad range of laser fluences and external field values allowed us to obtain a set of traces of the magnetization dynamics. Figure 8 shows a great deal of similarity between simulations and experimental data (cf. Fig. 3). It is seen that the frequency increases with increasing external field $H$ while the amplitude decreases (see Fig. 8a). The simulations for various stimulus strengths show the expected growth of the precession amplitude (see Fig. 8b).

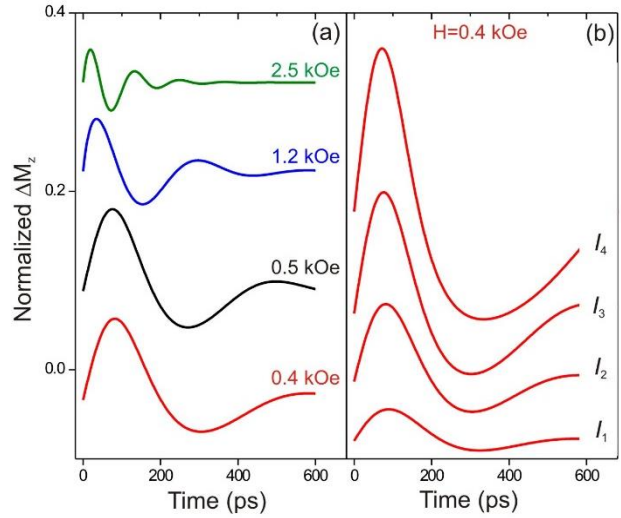

**Fig. 8.** Photo-magnetic precession obtained in numerical simulations of the LLG equation for: a) field dependence at moderate excitation level and b) power dependence (I=4, 10, 16, and 22 arb. units) at $H = 0.4$ kOe.

We further repeated our fit procedure with Eq.(1) to obtain the precession parameters from these data. Figure 9 shows the values of the amplitude and frequency of the precession in the power regime. At a low field $H = 0.4$ kOe (red), the nonlinearity is clearly visible and comparable with experimental data, as seen in Fig. 4. Similarly, at high fields (green), the behavior is mostly linear. Figure 9a shows a great deal of similarity between simulations (amplitude parameter) and experimental data (normalized value $A_F/\theta_{max}$) (cf. Fig. 4a). The analysis of the damping parameter (Fig. 9c) also confirms the experimental findings (as in Fig. 7a), revealing the existence of two regimes, linear and nonlinear. The results of the simulations confirm that the observation of the nonlinear response of the magnetic system can be attributed to the anharmonicity of the energy landscape.

Notably, in the simulations, as well as in the experimental data, we not only observe a second-order correction to the effective damping $\alpha_2$, but also a deviation from Eq.(7) at even larger amplitudes (cf. Fig. 7a and Fig. 9c). The latter manifests as a reduction of the effective damping compared to the expected $\alpha_0 + \alpha_2 A^2$ dependence shown with dashed lines. This higher-order effect is unlikely to originate in the multi-magnon scattering contribution since the latter would only further increase the effective damping [8]. We rather believe that this is likely an artifact of the used damped oscillator model where in the range of $\alpha_{eff} \approx 1$ the quasiperiodic description of magnetization precession ceases to be physically justified.

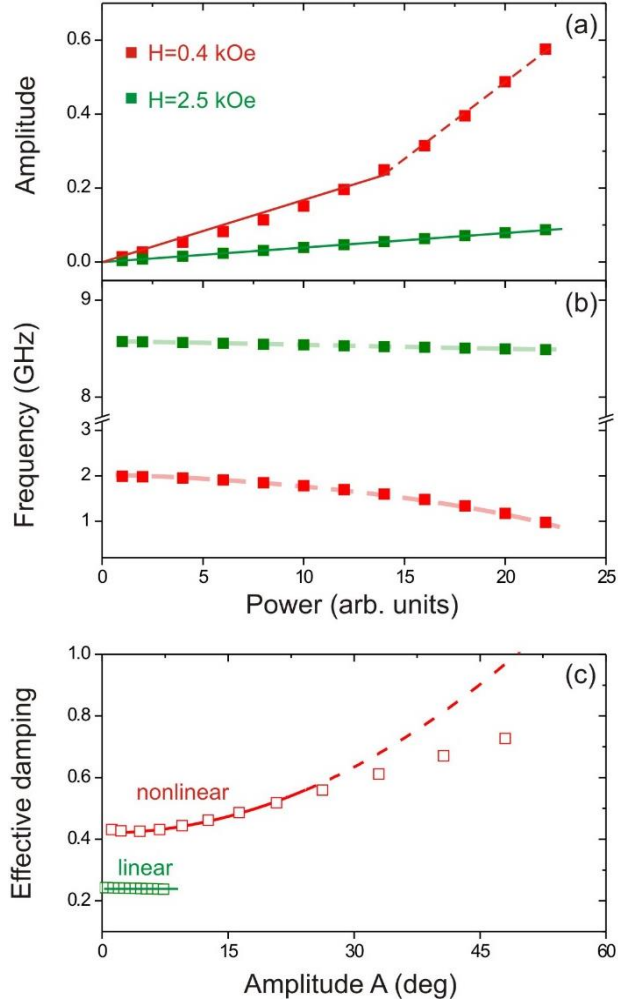

**Fig. 9.** Power dependence of the a) amplitude and b) frequency as obtained in the simulations for low (red dataset) and high (green dataset) external magnetic fields. c) Effective damping in the linear and nonlinear precession regimes.

## 6. Photo-induced phase diagram of sub-switching regime

It is seen from both experimental and numerical results above that the cubic symmetry of the magnetic system is key for the observed nonlinear magnetization dynamics. To quantify the parameter space for the nonlinearity, we first estimate the realistic values of the effective light-induced magnetic field $H_L$. Throughout a number of works on photo-magnetism in Co-doped garnets, a single-ion approach to magnetic anisotropy is consistently utilized. We note that in YIG:Co, it is the Co ions at tetrahedral sites that are predominantly responsible for the cubic anisotropy of the magnetic energy landscape [22]. In the near-IR range, these ions are resonantly excited at the 1300 nm wavelength, resulting in improved efficiency of the photo-magnetic stimulus, as compared to previous works [45]. Further, we note that at the magnetization switching threshold, about 90% of the $Co^{3+}$ ions with a concentration on the order of $10^{20}$ cm$^{-3}$ are excited with incident photons [11, 46]. Taking into account the single-ion contribution to the anisotropy $\Delta K_1 \sim 10^5$ erg/cm$^3$ [47], and assuming a linear relation between the absorbed laser power (or fluence) and the effective photo-magnetic field $H_L$, for the latter we find that $H_L \sim 1$ kOe is sufficient for the magnetization switching. This means that the sub-switching regime of magnetization dynamics (cf. Fig. 1) refers to the laser fluences (as well as wavelengths), resulting in smaller effective fields.

We reiterate that in previous works, the impact of the external magnetic field on the photo-magnetically driven magnetization precession has not been given detailed attention. To address this gap, we plotted the amplitude of the precession $A$ calculated in the same way as above in the sub-switching regime (Fig. 10). As expected, the amplitude generally increases with $H_L$. However, we note a critical external field of about 0.5 kOe at which the desired amplitudes can be reached at smaller light-induced effective fields $H_L$. At this field, where the system enters a single domain state, the potential curvature around the energy minimum decreases, thus facilitating the large-angle precession. In other words, external magnetic fields can act as leverage for the effective field of the photo-induced anisotropy, thus reducing the magnetization switching threshold. An exhaustive study of magnetization switching across the parameter space shown in Fig. 10 remains an attractive perspective for future studies.

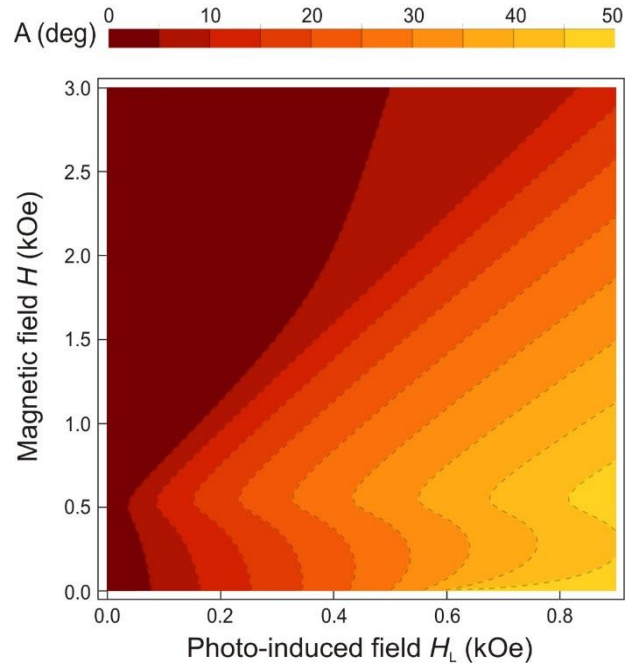

**Fig. 10.** Calculated amplitude map of the photo-induced magnetization precession in YIG:Co film.

In our analysis, we only considered a truly photo-magnetic excitation and neglected the laser-induced effects of thermal origin. It is, however, known that laser-driven heating can introduce an additional, long-lasting modification of magnetic anisotropy in iron garnets [48, 49]. The relatively long relaxation times associated with cooling are responsible for the concomitant modulation of the precession parameters and thus facilitate nonlinearities in the response of the magnetic system. Yet, 1300 nm laser excitation of magnetization dynamics in YIG:Co film was shown to be highly polarization-dependent [6], thus indicating the dominant role of the non-thermal excitation mechanism. On the other hand, the unavoidable laser-induced heating with experimental values of laser fluence in YIG:Co film has been estimated to not exceed 1 K [6]. As such, we do not expect modification of the Gilbert damping associated with the proximity of the magnetization compensation or Néel temperature in the ferromagnetic garnet [50]. However, a detailed investigation of the temperature-dependent nonlinear magnetization dynamics in the vicinity of the compensation point or a magnetic phase transition [51, 52] represents another promising research direction. Further, exploring the nonlinear regime in the response of the magnetic system to intense THz stimuli along the lines discussed in [33] enjoys a rich potential for spintronic applications.

## 7. CONCLUSIONS

In summary, we studied, both experimentally and numerically, the nonlinear regime of magnetization dynamics in photo-magnetic Co-doped YIG film. After excitation with femtosecond laser pulses at fluences below the magnetization switching threshold, there is a range of external magnetic field where the magnetic system demonstrates strongly nonlinear precession characterized by a significant increase of the effective Gilbert damping. We attribute this nonlinearity to the anharmonicity of the potential for the magnetic oscillator enhanced by the dominant role of the cubic magnetocrystalline anisotropy. The effective damping and its nonlinear contribution, as obtained from numerical simulations, both demonstrate a very good agreement with the experimental findings. Simulations of the magnetization dynamics by means of the LLG equation further confirm the nonlinearity in the magnetic response below the switching limit. Finally, we provide estimations for the realistic, effective photo-magnetic fields $H_L$ and map the workspace of the parameters in the sub-switching, nonlinear regime of photo-induced magnetization dynamics.

**ACKNOWLEDGMENTS**

This work has been funded by the Foundation for Polish Science (Grant No. POIR.04.04.00-00-413C/17) and the National Science Centre Poland (Grant No. DEC-2017/25/B/ST3/01305).